\documentclass[12pt]{iopart}
\usepackage{iopams} 
\usepackage{graphicx} 
\usepackage{cite}
\usepackage{xspace}

\usepackage{color}

\newcommand\noi{\noindent}
\newcommand\be{\begin{equation}}
\newcommand\ee{\end{equation}}
\newcommand\tmyag{Tm$^{3+}$:YAG\xspace}

\begin{document}

\title{Securing coherence rephasing with a pair of adiabatic rapid passages}

\author{M F Pascual-Winter, R-C Tongning, T Chaneli\`ere and J-L Le Gou\"et}
\address{Laboratoire Aim\'e Cotton, CNRS-UPR 3321, Universit\'e Paris-Sud, B\^{a}timent 505, Campus Universitaire, 91405 Orsay Cedex, France.}
\ead{maria-florencia.pascual-winter@lac.u-psud.fr}

\begin{abstract}
Coherence rephasing is an essential step in quantum storage protocols that use echo-based strategies. We present a thorough analysis on how two adiabatic rapid passages (ARP) are able to rephase atomic coherences in an inhomogeneously broadened ensemble. We consider both the cases of optical and spin coherences, rephased by optical or radio-frequency (rf) ARPs, respectively. We show how a rephasing sequence consisting of two ARPs in a double-echo scheme is equivalent to the identity operator (\emph{any} state can be recovered), as long as certain conditions are fulfilled. Our mathematical treatment of the ARPs leads to a very simple geometrical interpretation within the Bloch sphere that permits a visual comprehension of the rephasing process. We also identify the conditions that ensure the rephasing, finding that the phase of the optical or rf ARP fields plays a key role in the capability of the sequence to preserve the phase of the superposition state. This settles a difference between optical and rf ARPs, since field phase control is not readily guaranteed in the former case. We also provide a quantitative comparison between $\pi$-pulse and ARP rephasing efficiencies, showing the superiority of the latter. We experimentally verify the conclusions of our analysis through rf ARP rephasing sequencies performed on the rare-earth ion-doped crystal Tm$^{3+}$:YAG, of interest in quantum memories.
\end{abstract}

\submitto{\NJP}

\maketitle

\section{Introduction}

Very recently, adiabatic rapid passages (ARP) have drawn the attention of the optical quantum storage community \cite{Lauro2011, Mieth2012}, as they offer the possibility of rephasing atomic coherences. In inhomogeneously broadened materials for quantum storage purposes, the fact that the atomic coherences dephase is the key property that allows the storage. At the same time, bringing the coherences back into phase is a necessity for retrieving the stored information (see, e.g. \cite{Tittel2010}). The coherences to be rephased can be both of spin or optical nature. Even though optical ones are needed to interact with the information-carrying photons, conversion to spin coherences is preferred in between the capture and the retrieval stages of the storage protocol in order to take profit of their longer lifetime. In such schemes, coherences of both types dephase and need to be rephased by optical and radio-frequency (rf) means \cite{Lovric2013}. This is the case of echo-based memories. Protocols based on electromagnetically induced transparency carefully eliminate the stage where the information is stored in the optical coherences and allow direct storage in the spin coherences \cite{Duan2001, Chaneliere2005} by means of optical control pulses. In this case, only rf rephasing is needed \cite{Mieth2012}. However, this technique is not best-suited for inhomogeneously broadened media due to its narrow-band characteristics. Neither matter the specific characteristics of the protocol nor the nature of the coherences, typically, these are rephased by the application of a $\pi$ pulse. However, disadvantages concerning $\pi$ pulses have been pointed out. One is the need of high powers for the optical or rf fields, which scales as the protocol bandwidth squared. Another one, concerning optical coherences only, is the unavoidable distortion of the $\pi$ pulse as it propagates through the sample because of light absorption \cite{Ruggiero2010}. This makes that the pulse area is no longer $\pi$ after some threshold depth in the sample. The third disadvantage, related to the previous one in the optical case, is the high sensitivity of the technique to spatial intensity variations. These drawbacks have made researchers explore ARPs as an alternative.

An ARP consists of a pulse whose frequency is chirped through a range that typically goes from much lower to much higher frequencies than the ones in the inhomogeneous width one wishes to rephase. The amplitude of the field can either be varied as well or kept constant (see figures \ref{fig:geometry}(a) and \ref{fig:geometry}(b)).

Adiabatic rapid passages have been extensively used in nuclear magnetic resonance (NMR) through decades. Initially, in the early 1960's, it proved to be a means for intensity-insensitive and frequency-selective adiabatic spin inversion \cite{Abragam1961}. Transfer to the optical domain came some years later in the experiments by Loy \cite{Loy1974, Loy1978}. A rich investigation on pulse shape optimization (amplitude and phase modulation), from analytically \cite{Allen1975, Hioe1984, Silver1985, Baum1983,Garwood1991, Vitanov1996, Tannus1997, Garwood2001} to numerically \cite{Ugurbil1988, Johnson1989, Garwood1991, Poon1995, Garwood2001} derived or proposed methods, opened the way to increase the intensity insensitivity and/or the frequency selectivity of the pulses. 

In 1987, Kunz approached the question of using ARPs for spin rephasing \cite{Kunz1987}. He highlighted the fact that rephasing is not possible with a regular ARP. Put in words more suitable for the present context of quantum information storage, the reason he stated is the following: the passage of the ARP, leaves the spins with a phase that is a function of the spin transition frequency of each particular atom. This phase needs to be compensated for in order to enable spin refocusing. From Kunz \cite{Kunz1987} and other research groups (see \cite{Bendall1987, Conolly1989} and references in \cite{Hwang1995}) raised a variety of more complex pulses, generally combining regular ARPs or half ARPs as building blocks, that provided phase compensation. However, rather early, Conolly and coworkers realized that just adding a second identical ARP is enough to compensate for the phase induced by the first, and they proved it experimentally \cite{Conolly1991}. Here, words need to be carefully chosen not to mislead the reader. The phase compensation occurs in the reference frame that rotates at the instantaneous frequency of the rephasing pulse field, and still further clarification is needed for the term \emph{phase compensation} to be exact since that frame changes abruptly from the first to the second ARP. In the laboratory frame, the term ``phase compensation'' is not appropriate. The picture is more complicated and we will discuss it in detail in the section that follows. In any case, coherences can be rephased indeed by the application of two ARPs. The pulses need not be consecutive, which allows for storage time. Rf realizations in the context of NMR can be found in refs. \cite{Schupp1993, deGraaf1995, Garwood2001}. More recently, interest in ARPs has risen in the quantum information community. Specific implementations for coherence rephasing have been demonstrated both in the rf \cite{Lauro2011, Mieth2012, JulsgaardArXiv2013} and optical \cite{Damon2011} domains in the context of broadband quantum memories. Nevertheless, recent applications of ARPs do not limit to their rephasing capabilities. Their use for optical-to-spin coherence transformation and back has been theoretically studied in ref. \cite{Minar2010}, showing that they preserve the collectivity of the superposition state of an atomic ensemble (i.e., the phases of each of the components involved in a Dicke state). Other kinds of adiabatic pulses, which satisfy the adiabatic condition (to be defined later) but which do not involve frequency chirps have also been developed. The Stark-chirped rapid adiabatic passage (SCRAP) compensates the difficulty of chirping short (nanosecond) optical pulses by inducing a time-varying shift in the atomic transition through a far off-resonance adiabatic optical pulse. Another time-delayed close-to-resonance adiabatic pulse performs the population transfer between two optical states \cite{Yatsenko1999, Rickes2000}. As another example, the stimulated Raman adiabatic passage (STIRAP) relies on the adiabatic variation of the amplitude of two detuned and time-delayed pulses to generate a population transfer between two spin states by optical means (see \cite{Vitanov2001} and references therein). 

Two articles provide enlightening analysis of how rephasing by two ARPs happens. These are refs. \cite{Hwang1995}, by Hwang and Shaka, and \cite{deGraaf1997}, by de Graaf and Nicolay. The latter authors use the most spread out approach in NMR, that is, 3D geometrical representations of the trajectory of the effective magnetic field (the \emph{control vector}, in a language better suited to the present article) and that of the magnetization vector (the \emph{Bloch vector}) under the effect of the former. Hwang and Shaka proposed a much more concise approach by choosing a matrix treatment of the effect of each ARP. Then, a rephasing or any other sequence is constructed as the product of the matrices corresponding to the building blocks. This way, conclusions can be drawn very easily. However, Hwang an Shaka limited their analysis to one particular type of ARP-based sequences, of interest in NMR (intended to eliminate the deleterious effects of the huge water resonance in the quality of NMR spectra). 

In this article, we propose a matrix treatment of the rephasing sequence involving two ARPs, in the Bloch sphere formalism. We will revisit and extend the analysis of the matrix computation for one ARP only, already undertaken in a previous work by some of the authors of the present one \cite{Lauro2011}. Later, we will make use of the ARP matrix to build up the matrix associated to a rephasing sequence. We will see that this matrix approach allows a simple understanding of how the rephasing works and why two ARPs are needed to rephase an inhomogeneously broadened ensemble of atomic coherences. A geometrical interpretation  derived from the matrix treatment provides a visual comprehension of the process. We will draw comparisons with $\pi$ pulses and regular $\pi$-pulse-based echo sequences in terms of rephasing efficiencies. We will also analyse the role of the rephasing field phase, as well as the phase preservation characteristics of the ARP-based rephasing sequence, particularly relevant for quantum memories. At this point, a significant distinction between rf and optical rephasing will be highlighted, originated in the different capabilities each technology provides for controlling the (rf or optical) field phase. We will present experimental verification of the critical conditions that enable rephasing. Radio-frequency spin rephasing experiments are performed on a rare-earth ion-doped crystal, a thulium-doped YAG, a material that has been actively studied as a candidate for quantum memories \cite{deSeze2006, Louchet2007, Louchet2008, Lauro2009, Lauro2009Jun, Chaneliere2010, Bonarota2010, Bonarota2011, PascualWinter2012}.

The paper is organised as follows: In a theoretical section we will deal with the derivation of the matrix expression for an ARP from the Bloch equations, the matrix construction of a two-ARP rephasing sequence, the comparison with $\pi$ pulses and considerations on phase preservation. The experiments are presented in the section coming after. Finally, some concluding remarks will be given.

\section{Theory}

The ARP matrix will be derived thanks to successive frame changes. The first one is well-known and uses the rotating frame within the rotating wave approximation (RWA). It allows an accurate and consistent relation between the excitation field phase and the atomic state phase. The second frame change takes the tipping control vector as a polar axis. It provides a simpler interpretation of the different angles in our geometric interpretation.

\subsection{One adiabatic rapid passage}


\subsubsection{The ARP matrix}\label{sec:ARPmatrix}

In this section we aim at obtaining a matrix expression for an ARP. We will disregard decoherence effects. In what follows, we will use the Bloch sphere representation for the dynamics of a two-level system, whose states will be denoted $\arrowvert a \rangle$ and $\arrowvert b \rangle$, in interaction with an external oscillating field $\bi{A}(\bi{r},t)$. The field can be either electric or magnetic, depending on the nature of the ARP, optical or rf, respectively. In such representation, the two-level system state is expressed by the Bloch vector $\bi{B}$ defined as

\be
\left[\bi{B}\right]_R  = \left(
\begin{array}{c}
\rho_{ab} + \rho_{ba} \\
\rmi (\rho_{ba}-\rho_{ab}) \\
\rho_{bb} - \rho_{aa}
\end{array}
\right),
\ee


\noi where $\rho_{ij}$ represents the element $\langle i \arrowvert \rho \arrowvert j \rangle$ of the density matrix operator $\rho$ and the notation $[\ ]_R$ indicates that the vector coordinates are given in the laboratory reference frame $R$ of cartesian axes $\hat{\bi{u}}$, $\hat{\bi{v}}$, $\hat{\bi{w}}$. We express the field as

\be
\bi{A}(\bi{r},t_0+t) = \bi{A}_0(t) \cos \left( \omega_0 t + \phi(t) \right),
\label{external field}
\ee

\noi where $\bi{A}_0(t)$ is a real vector, $\omega_0$ and $\phi(t)$ are real quantities and $t_0$ is the central instant of the ARP. In the Bloch sphere representation, the parameters that define the field compose the control vector in the following way

\be
\left[\tilde{\bbeta}(t_0+t)\right]_R = \left(
\begin{array}{c}
2 \Omega(t) \cos\left(\omega_0 t+\phi(t)\right) \\
0 \\
\omega_{ab}
\end{array}
\right),
\ee


\noi where $\Omega(t)$ is the Rabi angular frequency of the field and $\omega_{ab} = (E_b-E_a)/\hbar$ is the angular frequency of the transition between the states $\arrowvert a \rangle$ and $\arrowvert b \rangle$ of energies $E_a$ and $E_b$, respectively. In the case of an optical field, we have $\Omega(t) = \bmu_{ab} \cdot \bi{A}_0(t) / \hbar$, with $\bmu_{ab} = \langle a \arrowvert e \bi{r} \arrowvert b \rangle$ the electric dipole moment of the transition ($-e$ is the electron charge). We have chosen the relative phase of the states $\arrowvert a \rangle$ and $\arrowvert b \rangle$ so that $\bmu_{ab}$ is real. Thus, $\Omega$ is real as well. In the case of a magnetic field, we have $\Omega = \gamma B_1$, where $\gamma$ is the gyromagnetic factor and $B_1$ is the component of the oscillatory field perpendicular to the quantization axis.

In the Bloch sphere representation, Schr\"odinger's equation takes the simple form

\be
\dot{\bi{B}}(t) = \tilde{\bbeta}(t) \times \bi{B}(t),
\label{Blochequation}
\ee

\noi in the absence of decoherence. 
\Eref{Blochequation} describes the precession of $\bi{B}$ around $\tilde{\bbeta}$. The last statement might mislead the reader to think that the dynamics of $\bi{B}$ are rather simple. As a matter of fact, $\tilde{\bbeta}(t)$ is a vector that oscillates at frequency $\omega_0 + \dot{\phi}(t)$. Therefore, a vector precessing around it describes a complicated trajectory.

The picture becomes simpler if we change to a reference frame where the unit vector $\hat{\tilde{\bbeta}} = \tilde{\bbeta}/|\tilde{\bbeta}|$ is static. We will perform the transformation in two steps. First, we will change from the laboratory frame $R$ to the reference frame $R^\prime$ that rotates around the vertical axis at frequency $\omega_0 + \dot{\phi}(t)$ with phase $\phi(0)$ at $t=0$. $R^\prime$ is known as the rotating frame. The matrix

\be
C_1(t) = \left(
\begin{array}{ccc}
\cos(\omega_0 t+\phi(t)) & \sin(\omega_0 t+\phi(t)) & 0 \\
-\sin(\omega_0 t+\phi(t)) & \cos(\omega_0 t+\phi(t)) & 0 \\
0 & 0 & 1 \\
\end{array}
\right)
\label{matrixC1}
\ee

\noi performs the frame change from $R$ to $R^\prime$ through the operation $[\bi{B}(t_0+t)]_{R^\prime} = C_1(t) [\bi{B}(t_0+t)]_{R}$. This change of reference frame is equivalent to applying the transformation $\rho_{ab}(t_0+t) \mapsto \rho_{ab}^\prime(t_0+t) = \rho_{ab}(t_0+t) \exp(-\rmi(\omega_0 t+\phi(t)))$, $\rho_{ii} \mapsto \rho_{ii}^\prime = \rho_{ii}$ (with $i= a, b$) to the density matrix. As regards the control vector, its coordinates in the frame $R^\prime$ are given by


\be
\left[ \tilde{\bbeta}(t_0+t) \right]_{R^\prime} = \left(
\begin{array}{c}
\Omega(t) \left[ 1 + \cos(2(\omega_0 t +\phi(t))) \right] \\
-\Omega(t) \sin(2(\omega_0 t +\phi(t))) \\
\omega_{ab} 
\end{array}
\right),
\ee

\noi 
Taking the RWA, the above expression reduces to

\be
\left[ \tilde{\bi{\bbeta}}(t_0+t) \right]_{R^\prime} = \left(
\begin{array}{c}
\Omega(t) \\
0 \\
\omega_{ab} 
\end{array}
\right),
\ee

The Bloch vector dynamics in frame $R^\prime$ are ruled by

\begin{eqnarray}
\left[ \dot{\bi{B}}(t) \right]_{R^\prime} &=& \left[ \dot{\bi{B}}(t) \right]_{R} - (\omega_0+\dot{\phi}(t)) \hat{\bi{w}} \times \bi{B}(t) \nonumber \\
&=& \tilde{\bi{\bbeta}} (t) \times \bi{B}(t)  - (\omega_0+\dot{\phi}(t)) \hat{\bi{w}}^\prime \times \bi{B}(t)  \nonumber \\
&=& \bi{\bbeta} (t)  \times \bi{B}(t) 
\label{BlochequationRprime}
\end{eqnarray}

\noi with a new control vector $\bi{\bbeta}$ defined as

\be
\left[ \bi{\bbeta}(t_0+t) \right]_{R^\prime} = \left(
\begin{array}{c}
\Omega(t) \\
0 \\
\Delta - \dot{\phi}(t)
\end{array}
\right),
\ee

\noi and $\Delta = \omega_{ab} - \omega_0$. The dynamics of $\bi{B}$ in frame $R^\prime$ given by \eref{BlochequationRprime} are much simpler than in frame $R$. Indeed, in $R^\prime$ and under the RWA, the fast oscillatory behaviour of the control vector has been ruled out. What is left is just its smooth variation as $\Omega(t)$ and $\dot{\phi}(t)$ evolve during the ARP. The smoothness of that variation is ensured by the adiabatic condition to be described shortly. The Bloch vector now precesses around the slowly varying axis $\hat{\bbeta}(t)$ with angular frequency $\vert \bbeta(t) \vert$.

\begin{figure}
\begin{center}
\includegraphics[width=0.25\textwidth,angle=-90]{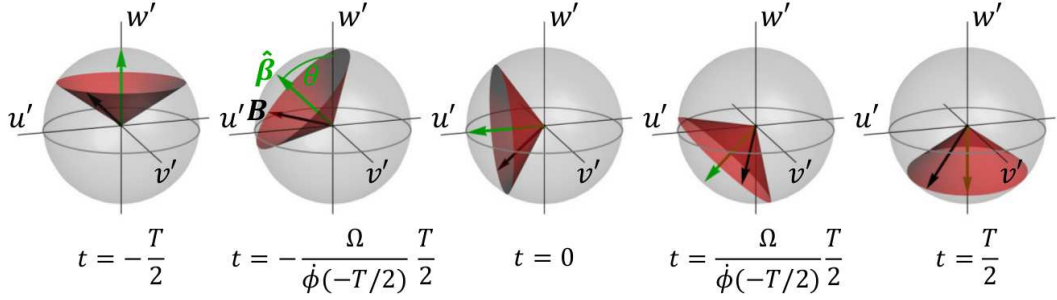}
\caption{\label{ARPsimple} (Colour online) Bloch sphere scheme of an ARP satisfying $\Delta = 0$ and $\vert \Omega(\pm T/2)/\dot{\phi}(\pm T/2)\vert \ll 1$. We have chosen $\dot{\phi}(t) = rt$ with $r>0$ and $\Omega(t)=\Omega$. At $t=-T/2$, the control vector $\bbeta$ points upwards. The control vector slowly starts turning upside down. During the lapse $-\Omega/(2 \dot{\phi}(-T/2)) T \lesssim t \lesssim \Omega/(2 \dot{\phi}(-T/2)) T$, the control vector stays off the vertical axis. This is a very small fraction of $T$ given the assumption mentioned above. Halfway through the passage, the control vector is in the equator. At the end of the passage, the control vector points downwards. All throughout the passage, the Bloch vector precesses around the control vector, describing the cones shown in the figure. On each snapshot, the Bloch vector is represented at an arbitrary position within the cone.}
\end{center}  
\end{figure}

In an ARP, the frequency of the external field is varied through a wide range, from values much smaller than $\omega_{ab}$ to  values much larger than $\omega_{ab}$ (or inversely, depending on the sign of $\ddot{\phi}$). In an ideal positively chirped ARP, the instantaneous detuning of the external field from the atomic frequency, $\Delta-\dot{\phi}(t)$, varies from $\infty$ to $-\infty$. At the same time, $\Omega(t)$ stays bounded, i.e. $\vert \Omega(t) \vert \ll \infty$. Therefore, in the course of the ARP,  the unit vector $\hat{\bbeta}(t_0+t)$ somehow goes from $\hat{\bi{w}}^\prime$ to $-\hat{\bi{w}}^\prime$, with $\hat{\bi{w}}^\prime$ the vertical axis of frame $R^\prime$ ($\hat{\bi{w}}^\prime \equiv \hat{\bi{w}}$). The exact trajectory and instantaneous angular velocity of $[\hat{\bbeta}(t_0+t)]_{R^\prime}$ will depend on the specific way $\Omega(t)$ and $\dot{\phi}(t)$ are varied during the ARP, as well as on the detuning $\Delta$. The simplest case, occurring for a constant $\Omega$, a linearly chirped frequency ($\phi(t) = rt$) and zero detuning halfway through the ARP ($\Delta = 0$) is depicted in \fref{ARPsimple}. The precession of the Bloch vector is also represented.

We mentioned above that a second change of reference frame is useful to further simplify the trajectory of the Bloch vector. In this case, we will change from $R^\prime$ to a reference frame $R^{\prime\prime}$ where the vertical axis $\hat{\bi{w}}^{\prime\prime}$ is instantaneously parallel to $\bbeta(t_0+t)$. This can be achieved through the matrix

\be
C_2(t) = \left(
\begin{array}{ccc}
\cos\theta(t) & 0 & -\sin\theta(t) \\
0               & 1 & 0                      \\
\sin\theta(t)  & 0 & \cos\theta(t) 
\end{array}
\right),
\ee

\noi which ensures $[\bi{B}(t_0+t)]_{R^{\prime\prime}} = C_2(t) [\bi{B}(t_0+t)]_{R^\prime}$. The angle $\theta(t)$ is defined as

\be
\fl\cos\theta(t) = \frac{\Delta - \dot{\phi}(t)}{\left[ \Omega(t)^2+(\Delta - \dot{\phi}(t))^2 \right]^{1/2}} \quad , \quad \sin\theta(t) = \frac{\Omega(t)}{\left[ \Omega(t)^2+(\Delta - \dot{\phi}(t))^2 \right]^{1/2}}.
\ee

\noi and is sketched in \fref{ARPsimple}. In $R^{\prime\prime}$, $\bbeta(t_0+t)$ takes the form

\be
[\bbeta(t_0+t)]_{R^{\prime\prime}} = \left(
\begin{array}{c}
0 \\ 0 \\ \left[ \Omega(t)^2+(\Delta - \dot{\phi}(t))^2 \right]^{1/2}
\end{array}
\right).
\ee

\noi Thus $[\hat{\bbeta}]_{R^{\prime\prime}}$ is static. The dynamics now become

\begin{eqnarray}
\left[ \dot{\bi{B}}(t) \right]_{R^{\prime\prime}} &=& \left[ \dot{\bi{B}}(t) \right]_{R^{\prime}} - \dot{\theta}(t) \hat{\bi{v}}^\prime \times \bi{B}(t) \nonumber \\
&=& \left[ \bi{\bbeta}(t) - \dot{\theta}(t)\hat{\bi{v}}^\prime \right] \times \bi{B}(t)
\label{BlochequationRprimeprime}
\end{eqnarray}

\noi 
The dynamics simplify under the adiabatic approximation. It assumes that  $\bi{B}$ essentially precesses around $\bbeta$, i.e.

\be
\dot{\theta}(t)^2 \ll \bi{\bbeta}(t)^2 
\ee

\noi or, equivalently,

\be
\left| \frac{\dot{\Omega}(t) (\Delta-\dot{\phi}(t)) + \Omega(t) \ddot{\phi}(t)}{\left[ \Omega(t)^2+(\Delta-\dot{\phi}(t))^2 \right]^{3/2}}\right| < 1 
\label{adiabaticitycondition}
\ee

The adiabatic approximation assumes that the precession rate (generalized Rabi frequency) is much faster than the control vector tipping rate. The condition is usually found in literature expressed as

\be
\frac{\Omega^2}{r} > 1,
\label{simplifiedadiabaticitycondition}
\ee

\noi obtained for the worst-case scenario ($\Delta = 0$) occuring for a chirp such that $\dot{\Omega} = 0$ and $\ddot{\phi}(t) = r$ (linear chirp).

\noi If \eref{adiabaticitycondition} is valid, the dynamics reduce to

\be
\left[ \dot{\bi{B}}(t) \right]_{R^{\prime\prime}} = \bi{\bbeta}(t) \times \bi{B}(t)
\label{BlochequationRprimeprimeAdiab}
\ee

\noi Integration yields

\begin{eqnarray}
[\bi{B}(t_0+t_b)]_{R^{\prime\prime}} = U(t_b, t_a) [\bi{B}(t_0+t_a)]_{R^{\prime\prime}} \\
U(t_b, t_a) = \left(
\begin{array}{ccc}
\cos\chi(t_b, t_a) & -\sin\chi(t_b, t_a) & 0 \\
\sin\chi(t_b, t_a) & \cos\chi(t_b, t_a) & 0 \\
0 & 0 & 1
\end{array}
\right), \\
\chi(t_b, t_a) = \int_{t_a}^{t_b} \left[ \Omega(t^\prime)^2+(\Delta - \dot{\phi}(t^\prime))^2 \right]^{1/2} dt^\prime.
\label{matrixU}
\end{eqnarray}
where $\chi(t_b, t_a)$ represents the total precession angle

We have now gathered all the necessary elements to build a matrix expression for an ARP. The matrix associated to an ARP connecting the initial and the final states is

\be
M_\mathrm{ARP} = C_1(T/2)^{-1} C_2(T/2)^{-1} U(T/2, -T/2) C_2(-T/2) C_1(-T/2),
\label{ARPmatrix}
\ee

\noi with $T$ the duration of the ARP. Due to the inclusions of $C_1$ and $C_2$, $M_\mathrm{ARP}$ acts on the reference frame $R$. It is worth noting that $C_1$ and $C_2$ are only necessary to be known at the instants of the beginning and end of the ARP.

\subsubsection{Far off-resonance initial and final conditions}

As we have said above, an ideal ARP is the one where $\bbeta$ departs from $\hat{\bi{w}}^\prime$ at $t_0-T/2$ and reaches $-\hat{\bi{w}}^\prime$ at $t_0+T/2$, in the case of a positive chirp (from lower to higher frequencies). We can simply understand that by analysing the case of an ARP-driven population inversion experiment. The Bloch vector is initially parallel to $\pm \hat{\bi{w}}^\prime$. The ARP will most efficiently drive the Bloch vector all the way to the $\mp \hat{\bi{w}}^\prime$ direction only if $\bbeta$ is initially parallel (or anti-parallel) to $\bi{B}$ and remains so all throughout the passage. In practical realizations, this condition is most closely satisfied if

\be
\left| \frac{\Omega(\pm T/2)}{\Delta-\dot{\phi}(\pm T/2)} \right| \ll 1. 
\label{condicion180}
\ee

\noi The detuning should be larger than the Rabi frequency or, in other words, the initial and final excitations should be far off-resonance. 
The expression has to be satisfied for any atomic detuning $\Delta$  within the inhomogeneous broadening $\Gamma_\mathrm{inh}$.

If \eref{condicion180} is satisfied, the matrices $C_2(-T/2)$ and $C_2(T/2)$ reduce to

\be
C_2\left(-\frac{T}{2}\right) = \mathbb{I}
\quad , \quad
C_2\left(\frac{T}{2}\right) = \left(
\begin{array}{ccc}
-1 & 0 & 0  \\
0 & 1 & 0  \\
0 & 0 & -1 
\end{array}
\right)
\label{matrixC2}
\ee

\noi Therefore, the mathematical treatment of the ARP matrix \eref{ARPmatrix} will simplify significantly.

In what follows we will assume \eref{condicion180} is valid.

\subsubsection{Eigensystem}\label{sec:eigensystem}

In \sref{sec:ARPmatrix} we have deduced an analytical expression for the ARP matrix. However, we still do not have a feeling of what the effect of $M_\mathrm{ARP}$ is when applied to an arbitrary Bloch vector. For getting it, we will calculate its eigensystem $M_\mathrm{ARP} \bi{a}_i= \lambda_i \bi{a}_i $. The eigenvalues and eigenvectors are

\be
\fl\begin{array}{lllll}
\lambda_1 = -1 & , & \lambda_2 = -1 & , & \lambda_3 = 1 \\
\bi{a}_1 = \left( \begin{array}{c} 0 \\ 0 \\ 1 \end{array} \right) & , &
\bi{a}_2(\Delta) = \left( \begin{array}{c} \cos(\varphi(\Delta)) \\ \sin(\varphi(\Delta)) \\ 0 \end{array} \right) & , &
\bi{a}_3(\Delta) = \left( \begin{array}{c} \cos(\varphi(\Delta) + \frac{\pi}{2}) \\ \sin(\varphi(\Delta) + \frac{\pi}{2}) \\ 0 \end{array} \right) 
\end{array},
\label{eigensystem}
\ee

\noi with

\be
\varphi(\Delta) = \frac{1}{2} \left[ -\chi(\Delta) + \phi\left(-\frac{T}{2} \right) + \phi\left( \frac{T}{2} \right) \right].
\label{angulo}
\ee

\noi We see from the set of eigenvalues that $M_\mathrm{ARP}$ is a rotation matrix of angle $\pi$. The rotation axis is $\bi{a}_3$. This is the first relevant conclusion of our analysis: the effect of an ARP on a Bloch vector is equivalent to that of a $\pi$ rotation about an axis contained in the equatorial plane of the Bloch sphere. This by no way means that the Bloch vector actually \emph{performs} this rotation. As we have discussed above, in the frame $R^\prime$, the Bloch vector precesses around the control vector as this one goes from $\hat{\bi{w}}^\prime$ to $-\hat{\bi{w}}^\prime$. To get the motion in the frame $R$, we still need to compose that motion with the rotation about $\hat{\bi{w}}$ involved in the $R$ to $R^\prime$ frame change. Indeed, the trajectory of the Bloch vector is much more complex than a $\pi$ rotation. However, if we take snapshots of the Bloch vector right before and right after an ARP, a $\pi$ rotation about an axis contained in the equatorial plane of the Bloch sphere links the two pictures.

It is important to note that the rotation axis not only depends on parameters of the ARP pulse, such as $\phi(\pm T/2)$ or the time profile of $\Omega(t)$ and $\dot{\phi}(t)$. It also depends on the transition frequency of the two-level system through $\chi$. We have highlighted this dependence by stating explicitly in \eref{angulo} that $\chi$ and, hence, $\varphi$ and $\bi{a}_3$ are functions of $\Delta$. This becomes relevant when dealing with an inhomogeneously broadened ensemble. In such a case, each frequency class experiences a $\pi$ rotation about a different axis, although all of these axes are contained in the equatorial plane of the Bloch sphere. For a given frequency class, the specific orientation of the rotation axis is determined by $\chi(\Delta)$.

\subsubsection{Comparison with a $\pi$ pulse}\label{sec:CompPiPulse}

The ARP feature of being equivalent to a $\pi$ rotation about an axis contained in the equatorial plane reminds us of a $\pi$ pulse. In the latter case, if the external field is perfectly tuned to the two-level system, the Bloch vector performs a $\pi$ rotation about an axis contained in the equatorial plane. The differences between an ARP and a perfectly tuned $\pi$ pulse are, first, the rotation axes (except if the parameters for the ARP are especially chosen), and second, the fact that the Bloch vector \emph{does} perform the rotation in the case of the $\pi$ pulse. 

Aside from that, the main dissimilarity between an ARP and a $\pi$ pulse rises when it comes to consider a spectral distribution of two-level systems. The dependence of these pulses on the detuning is quite different. In the case of an ARP, we can see from \eref{eigensystem} that the rotation axes for a range of $\Delta$ values span out on the equatorial plane. Nonetheless, the rotation angle is $\pi$ for all the two-level systems, despite their detuning. On the other hand, in the case of a $\pi$ pulse, the rotation axes for a range of $\Delta$ values fan out on a vertical plane, let us say, the plane $u^\prime w^\prime$. Moreover, the rotation angle is not $\pi$ for everybody. It is rather given by the expression $\int (\Omega(t)^2+\Delta^2)^{1/2} dt$, that reduces to $[1+(\Delta/\Omega)^2]^{1/2} \pi$ for a square pulse. The point is to analyse how one and the other kind of dependence on $\Delta$ impacts on the result of a sequence containing either an ARP or a $\pi$ pulse.

\paragraph{Photon or spin echo:\\}

One of the most current uses of a $\pi$ pulse is the photon or spin echo, where the point is to rephase an inhomogeneous distribution of two-level systems. Starting from a set of Bloch vectors oriented along, let us say, the $\hat{\bi{v}}^\prime$ direction, the echo sequence, consisting in a waiting time $\tau$, a $\pi$ pulse and another waiting time $\tau$ (symbolized $\tau$-$\pi$-$\tau$), is ideally intended to yield $-\hat{\bi{v}}^\prime$-oriented Bloch vectors. For the application of such a sequence to a detuned two-level system, we will define the rephasing error $\varepsilon_\pi$ as the angle between the equatorial component of the final Bloch vector for the case $\Delta/\Omega \ll 1$ and the final Bloch vector of the atom in resonance with the field. It can be shown that $|\varepsilon_\pi|$ scales as $(\Delta/\Omega)^2$ (the exact expression for a square pulse is $\varepsilon_\pi = \sin(2 \Delta \tau) (\Delta/\Omega)^2$). 

Due to the similarities between the ARP and the $\pi$ pulse, it is valid to ask oneself whether the former can be used as a rephasing pulse in an echo experiment. Rephasing, if it occurs, will not necessarily happen with the Bloch vectors parallel to $-\bi{v}^\prime$. The direction of the rephased Bloch vectors will rather be the one the resonant vector takes at the end of the sequence. In any case, simple arguments immediately tell us that an ARP in an echo sequence does not seem a clever choice: As the orientation of the Bloch vector in the equatorial plane is a crucial issue in the rephasing experiment, an axial operation whose axis is variably oriented in the equatorial plane, as an ARP, will evidently not work. Let us now consider quantitative arguments. For the sequence $\tau$-ARP-$\tau$, we define $\varepsilon_\mathrm{ARP}$ in the same way as we did for a $\pi$ pulse. Simple calculations yield that $\varepsilon_\mathrm{ARP}$ scales as $\Delta^2/r$, where $r$ is the mean value of the chirp rate $\ddot{\phi}(t)$ (the equality $\varepsilon_\mathrm{ARP} = \Delta^2/r$ is obtained for a linearly chirped ARP). Due to the  adiabatic condition \eref{simplifiedadiabaticitycondition}, we deduce that $|\varepsilon_\mathrm{ARP}| \gg (\Delta/\Omega)^2 \sim |\varepsilon_\pi|$. We conclude that an ARP is unsuited for substituting a $\pi$ pulse in a regular echo sequence. At least, it will be much less efficient than the standard $\pi$ pulse.

\paragraph{Population inversion:\\}

Another purpose for which $\pi$ pulses and ARPs are often used is population inversion. Let us compare the performances of both pulses. In this case, we will define $\varepsilon$ as the angle between the final Bloch vector and the desired vertical direction. For an ARP we have $\varepsilon_\mathrm{ARP}=0$ since, as we have seen, the ARP can be viewed as a $\pi$ rotation about an horizontal axis as long as \eref{condicion180} and \eref{adiabaticitycondition} are satisfied. As regards the $\pi$ pulse, for small $\Delta/\Omega$ we get $\varepsilon_\pi \simeq 2 \Delta/\Omega$. It is clear, then, that the choice for population inversion purposes will be an ARP rather than a $\pi$ pulse.

\subsection{Rephasing by two adiabatic rapid passages}

\subsubsection{The rephasing sequence matrix}\label{sec:repasingseqmat}

As we have seen in the previous section, an ARP is not able to rephase an inhomogeneously broadened ensemble of two-level systems unless $\Gamma_\mathrm{inh}$ is small enough or the chirp is fast enough to fulfil $\Gamma_\mathrm{inh}^2/r \ll 1$. It is known \cite{Conolly1991, Garwood2001, Lauro2011, Mieth2012}, however, that two ARPs can produce the rephasing without any stringent condition other that the fulfuilment of \eref{adiabaticitycondition} and \eref{condicion180} for the individual ARPs. In this section, we will analyse how the rephasing by two ARPs happens.

Once again, we prefer to make use of the matrix representation of the process. The sequence to be considered is the following: the coherences are created at $t=0$, the system evolves freely during a time $\tau_1$, a first ARP (henceforth called ARP A) is applied between $t_\mathrm{A} - T_\mathrm{A}/2 = \tau_1$ and $t_\mathrm{A} + T_\mathrm{A}/2$, the system evolves freely again during a time $\tau_2$, a second ARP (henceforth called ARP B) is applied between $t_\mathrm{B} - T_\mathrm{B}/2 = \tau_1 + \tau_2 + T_\mathrm{A}$ and $t_\mathrm{B} + T_\mathrm{B}/2$, the system evolves freely during a time $\tau_3$. We read the coherence at instant $t= \tau_1+\tau_2+\tau_3+T_A+T_B$. The matrix associated to the sequence is

\be
L = F(\tau_3) M_\mathrm{ARPB} F(\tau_2) M_\mathrm{ARPA} F(\tau_1),
\label{matrixT}
\ee

\noi with 

\be
F(\tau) = \left(
\begin{array}{ccc}
\cos(\omega_{ab} \tau) & -\sin(\omega_{ab} \tau) & 0 \\
\sin(\omega_{ab} \tau)  &  \cos(\omega_{ab} \tau) & 0 \\
0 & 0 & 1
\end{array}
\right),
\ee

\noi the matrix for the free evolution time $\tau$ in frame $R$, and 

\be
\fl M_{\mathrm{ARP}J} = \left[C_{1,J}\left(T_J/2\right)\right]^{-1} \left[C_{2}\left(T_J/2\right)\right]^{-1} U_J\left(T_J/2,-T_J/2\right) C_{2}\left(-T_J/2\right) C_{1,J}\left(-T_J/2\right),
\ee

\noi the matrix for the ARP $J=\mathrm{A},\mathrm{B}$. We have included de subindex $J$ in $C_1$ to indicate that for each ARP, the matrix in \eref{matrixC1} is to be calculated using the parameters $\omega_{0}$ and $\phi$ of ARP $J$, henceforth noted $\omega_{0,J}$ and $\phi_J$ (subindexation is made extensive to all other ARP-dependent quantities). As for $C_2$, the subindex is unnecessary since we consider that \eref{matrixC2} is valid for both ARPs.

The matrix computation of $L$ in \eref{matrixT} yields

\be
L = \left(
\begin{array}{ccc}
\cos\alpha & -\sin\alpha & 0 \\
\sin\alpha & \cos\alpha & 0 \\
0 & 0 & 1
\end{array}
\right),
\label{matrixL}
\ee

\noi with

\begin{eqnarray}
\alpha &=& \omega_{ab} (\tau_1-\tau_2+\tau_3) - \chi_\mathrm{B}(\Delta_\mathrm{B}) + \chi_\mathrm{A}(\Delta_\mathrm{A}) + \phi_\mathrm{B}(-T_\mathrm{B}/2) + \phi_\mathrm{B}(T_\mathrm{B}/2) \nonumber \\ 
&&  - \phi_\mathrm{A}(-T_\mathrm{A}/2) - \phi_\mathrm{A}(T_\mathrm{A}/2). \nonumber \\
&=& \omega_{ab} (\tau_1-\tau_2+\tau_3) + 2 \left[ \varphi_\mathrm{B}(\Delta_\mathrm{B}) - \varphi_\mathrm{A}(\Delta_\mathrm{A}) \right]
\end{eqnarray}

\noi We see that $L$ is a counterclockwise rotation about $\hat{\bi{w}}$ of an angle that depends on parameters of the ARPs but also on the atom transition frequency. For the sequence to succeed in rephasing an inhomogeneous distribution, we need to get rid of the dependence on $\omega_{ab}$. This can be achieved by setting

\begin{eqnarray}
\tau_3 = \tau_2-\tau_1, \label{condition_tau3}\\
\chi_\mathrm{A}(\omega_{ab}-\omega_{0,\mathrm{A}}) = \chi_\mathrm{B}(\omega_{ab}-\omega_{0,\mathrm{B}}) \qquad \forall\ \omega_{ab} \in \Gamma_\mathrm{inh}, \label{condition_gamma}
\end{eqnarray}

\noi which leaves

\be
\alpha = \phi_\mathrm{B}(-T_\mathrm{B}/2) + \phi_\mathrm{B}(T_\mathrm{B}/2) - \phi_\mathrm{A}(-T_\mathrm{A}/2) - \phi_\mathrm{A}(T_\mathrm{A}/2).
\label{alphareduced}
\ee

\noi Fulfilling \eref{condition_tau3} is trivial: it is enough to set $\tau_2>\tau_1$ and to impose the rephasing instant a time $\tau_2-\tau_1$ after the second ARP. As for \eref{condition_gamma}, the simplest way to satisfy it is to use identical amplitude time profiles and identical frequency chirps for both ARPs. The phases need not be identical, though. Actually, it is the relative phase between ARPs both at the beginning and at the end of the pulses that determines the rotation angle of $L$ or, in other words, the phase of the rephased coherence. If the phases of the ARP fields are also equal, which makes the second ARP just a time-shifted version of the first, $L$ is the identity matrix. Another way of retrieving the identity matrix is to design the ARPs such that $\phi_J(T_J/2) + \phi_J(-T_J/2) = 2 n \pi$, with $n$ an integer.

\subsubsection{Phase preservation}

\begin{figure}
\begin{center}
\includegraphics[width=0.7\textwidth,angle=0]{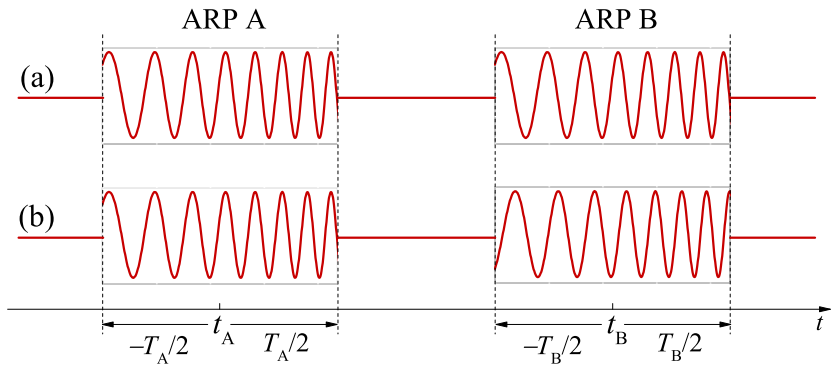}
\caption{\label{waves} (Colour online) Field profile for two linearly chirped ARPs of constant $\Omega$. The chirp rate, chirp range and field amplitude is the same for both ARPs. (a) Both ARPs have the same phase, i.e. $\phi_A(\pm T/2) = \phi_B(\pm T/2)$. This sequence gives phase preservation rephasing. (b) The ARPs have different phases, i.e. $\phi_A(\pm T/2) \neq \phi_B(\pm T/2)$. This sequences adds a phase factor to the rephased coherences.}
\end{center}  
\end{figure}

The above point about relative phases introduces a difference between rf and optical rephasing. Current rf technology permits accurate and arbitrary control not only of the amplitude but also of the phase of the rf field. The production of two identical pulses, identical both in their envelope and carrier, as depicted in \fref{waves}(a), is straightforward. Spin rephasing with such ARPs preserves the phase of the initial superposition state ($L=\mathbb{I}$). On the other hand, the phase of an optical field is trickier to control. If the optical ARP pulses are to be produced from acousto-optical modulation of a monochromatic cw laser, the acousto-optic modulator allows one to control the relative phase between pulses, as long as the laser field stays coherent through the rephasing sequence. So the same situation as for the rf case is found. If the laser coherence time is shorter than the rephasing sequence, no phase control can be applied and the phase of the rephased coherence will be random. This is the situation sketched in \fref{waves}(b). 

For the processing or storage of quantum information encoded in the phase of the carrier photons (i.e. time-bin qubits), the above considerations become relevant. 

Having said that, we must prevent the reader from thinking that phase randomization can be avoided by preferring storage in the spin coherences. Acces to those coherences, either by conversion from optical ones or by EIT, involves the application of additional external light pulses. The random phase of those pulses results in phase randomization of the retrieved qubit as well.

\subsubsection{Geometric interpretation}

\begin{figure}
\begin{center}
\includegraphics[height=1.0\textwidth,angle=-90]{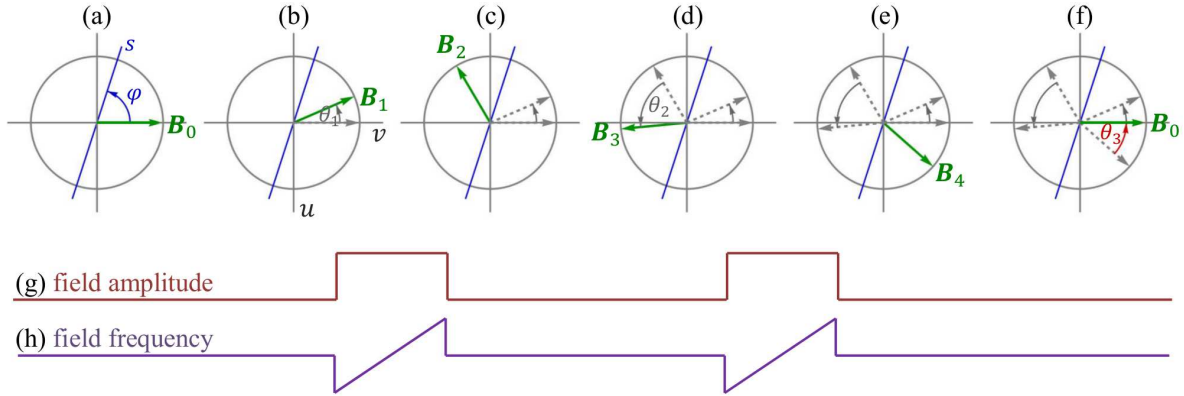}
\caption{\label{fig:geometry} (Colour online) (a-f) 2D geometrical representation of rephasing by two identical ARPs. The plane $uv$ is the equator of the Bloch sphere. The initial vector $\bi{B}_0$ (a) describes a first rotation of angle $\theta_1 = \omega_{ab}\tau_1$ during the first free evolution period (b). Then, the first ARP applies a reflection in axis $s$ leading from $\bi{B}_1$ to $\bi{B}_2$ (c). A second rotation of angle $\theta_2 = \omega_{ab}\tau_2$ follows, corresponding to the second free evolution period (d). The Bloch vector gains position $\bi{B}_3$. From there, the second ARP takes it to $\bi{B}_4$ by means of a reflection in axis $s$ (e). Finally, the last free evolution period applies a rotation of angle $\theta_3 = \omega_{ab}(\tau_2-\tau1)$ that leaves the Bloch vector at its initial position (f). (g,h) Amplitude (g) and frequency (h) time dependences of the rephasing field.}
\end{center}  
\end{figure}

The rephasing process by two ARPs is easily understood by means of a simple geometric picture. We will go through the steps of the rephasing sequence analysing their action on the Bloch sphere. We will content ourselves with a 2D picture of what happens in the equatorial plane. In other words, we will only consider the effect of these operations on the projection of the Bloch vector onto the $uv$ plane. As we concluded in \sref{sec:eigensystem}, the effect of one ARP can be mapped in the Bloch sphere to a $\pi$ rotation about an axis contained in the equatorial plane oriented at a counterclockwise (positive) angle $\varphi (\Delta) = 1/2 \left[ -\chi(\Delta) + \phi\left( -T/2 \right) + \phi\left( T/2 \right) \right] $ from the axis $\hat{\bi{v}}$ (see \eref{eigensystem} and \eref{angulo}). The 2D version of this operation  is a reflection in the mentioned axis. We will restrict to the case of two totally identical ARPs. Then, the rotation (3D) or reflection (2D) axis is the same for both ARPs. We will note this axis $s$, as shown in \fref{fig:geometry}. 

\Fref{fig:geometry} shows, with very simple elements, how rephasing by two ARPs works. We will start from a Bloch vector $\bi{B}_0$ initially oriented along $\hat{\bi{v}}$ (\fref{fig:geometry}(a)). From the geometrical point of view, the rephasing sequence consists in (see \fref{fig:geometry}):
\begin{itemize}
\item[(\emph{i})] \emph{The first free evolution period:} A rotation of $\theta_1=\omega_{ab}\tau_1$. This leads from $\bi{B}_0$ to $\bi{B}_1$ (\fref{fig:geometry}(b)).
\item[(\emph{ii})] \emph{The first ARP:} A reflection in axis $s$. This leads to $\bi{B}_2$ (\fref{fig:geometry}(c)).
\item[(\emph{iii})] \emph{The second free evolution period:} A rotation of $\theta_2=\omega_{ab}\tau_2$. This leads to $\bi{B}_3$ (\fref{fig:geometry}(d)).
\item[(\emph{iv})] \emph{The second ARP:} A second reflection in axis $s$. This leads to $\bi{B}_4$ (\fref{fig:geometry}(e)). 
\item[(\emph{v})] \emph{The third free evolution period:} A rotation of $\theta_3=\theta_2-\theta_1$. The Bloch vector regains its initial position $\bi{B}_0$ (\fref{fig:geometry}(f)).
\end{itemize}

\noi At instance (\emph{iv}), we see that, to recover the initial vector $\bi{B}_0$, all we need to do is to let $\bi{B}_4$ describe an angle $\theta_2-\theta_1$. That is equivalent to condition \eref{condition_tau3}, compulsory for the rephasing to take place, and is exactly what instance (\emph{v}) is about. 

In \fref{fig:geometry}, our choice of the angles $\varphi$, $\theta_1$ and $\theta_2$ has been totally arbitrary. Had we chosen a different set of angles, the result would have been the same. As $s$, $\theta_1$ and $\theta_2$ are determined by $\omega_{ab}$, this proves that every Bloch vector in the inhomogeneous broadening regains its position of departure. This guarantees rephasing. In other words, we see from \fref{fig:geometry} that $L|_{2D}$, the restriction of $L$ to the equatorial plane, satisfies  $L|_{2D} = \mathbb{I}_{2\times2}$.

Regarding the Bloch vector vertical component, which we have neglected so far, the two 3D $\pi$ rotations combine to leave it unaffected. Hence, $L = \mathbb{I}$.

If we consider ARPs with different phases (non-vanishing relative phase), the geometric interpretation becomes a little trickier because two different axes $s_\mathrm{A}$ and $s_\mathrm{B}$ are involved. Anyway, it is not hard to convince oneself that $L$ is a rotation about $\hat{\bi{w}}$ of angle $2(\varphi_\mathrm{B}-\varphi_\mathrm{A}) = \phi_\mathrm{B}(-T_\mathrm{B}/2) + \phi_\mathrm{B}(T_\mathrm{B}/2) - \phi_\mathrm{A}(-T_\mathrm{A}/2)  \nonumber 
- \phi_\mathrm{A}(T_\mathrm{A}/2)$, just as in \eref{alphareduced}.

\subsubsection{Comparison with $\pi$-pulse rephasing}

As we stated in \sref{sec:CompPiPulse}, the error associated to rephasing by a $\pi$ pulse scales as $(\Delta/\Omega)^2$. The same behaviour is found for $\varepsilon_{2\pi}$, the error corresponding to a rephasing sequence with two $\pi$ pulses, either of the same or opposite rotation axis (the errors for these two cases differ at higher order of $\Delta/\Omega$). In fact,

\be
\fl \varepsilon_{2\pi} = \left[ (\sin(2\Delta\tau_1) - \sin(2\Delta (\tau_2 - \tau_1)) - 2 \sin(\Delta \tau_2) + 2 \sin(\Delta (\tau_2 - 2 \tau_1))\right] \left(\frac{\Delta}{\Omega}\right)^2
\ee

\noi On the contrary, rephasing by two ARPs is $\Delta$-independent as long as \eref{condicion180}, \eref{adiabaticitycondition} and \eref{condition_gamma} are fulfilled. Therefore, its associated error, $\varepsilon_{2ARP}$, is zero.

For an experimental comparison of ARP- and $\pi$-pulse-based rephasing, see reference \cite{Mieth2012}.

\section{Experiments}

We have verified experimentally the rephasing time of the echo given by \eref{condition_tau3} and the phase preservation of the initial superposition state \eref{alphareduced} with two identical ARPs. 

The material system used was the rare-earth ion-doped crystal \tmyag (dopant concentration: $0.1$ at.\%), of interest in quantum storage applications \cite{deSeze2006, Louchet2007, Louchet2008, Lauro2009, Lauro2009Jun, Chaneliere2010, Bonarota2010, Bonarota2011, PascualWinter2012}. We have focused on rf ARP since accurate field control is available. A suitable rf spin transition is obtained in \tmyag by ground level splitting under the application of an external magnetic field of a fraction of Tesla. This transition is inhomogeneously broadened due to the slightly different lattice environment seen by each Tm$^{3+}$ impurity. A convenient optical transition at $793$ nm is used for initializing the system and probing its state. The technique used in the experiments was optically-detected NMR \cite{Kastler1950, Kastler1951, Brossel1952, Shelby1978}.

The optical aspects of the setup have been described extensively in \cite{Louchet2007} and \cite{Lauro2009Jun}. Basically, the light beam, emerging from an external cavity diode laser, is amplitude and phase-shaped by acousto-optic modulators, driven by a high sample-rate arbitrary wave form generator (AWG, Tektronix AWG5004). The crystal is cooled down to $1.7$ K in a liquid helium cryostat. The static magnetic field is generated by superconducting coils and oriented as in \cite{Louchet2007}. The rf ARP field is supplied by a $10$-turn, $20$-mm-long, $10$-mm-diameter coil oriented along the light pulse wave vector (the $[1\bar{1}0]$ axis of the cubic crystal lattice). The crystal sits at the coil centre. The rf signal, generated by the AWG, is fed to the coil through a $500$-W amplifier (TOMCO BT00500-AlphaSA) and a $\sim 600$-kHz-bandwidth resonant circuit. The amplitude of the static magnetic field ($\sim 0.5$ T) is chosen such that the spin transition frequency matches the rf circuit resonance known to be close to $14$ MHz. The inhomogeneous broadening of the spin transition in the sample was measured to be $\Gamma_\mathrm{inh} \sim 500$ kHz.

\begin{figure}
\begin{center}
\includegraphics[height=0.7\textwidth,angle=-90]{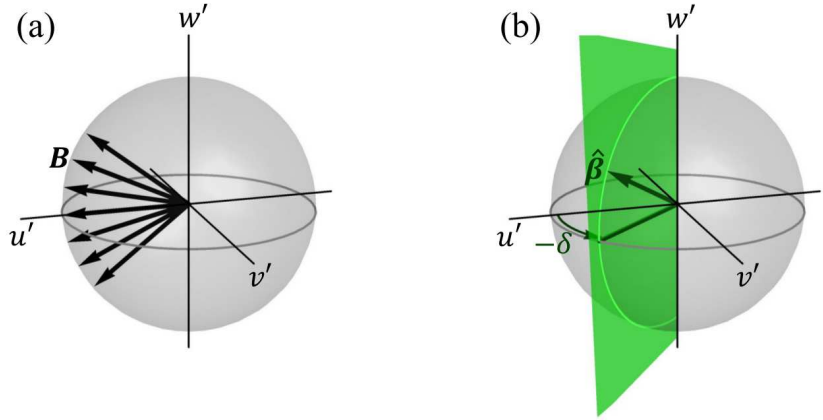}
\caption{\label{fig:AHP} (Colour online) (a) Span of Bloch vectors at the end of an AHP appied to an inhomogeneously broadened ensemble of atoms. The vector lying along $\hat{\bi{u}}^\prime$ corresponds to the atom in resonance with the last frequency of the AHP chirp (the central frequency of the interrupted ARP). The vectors in the upper ($[\Omega(0) \hat{\bi{u}}^\prime + (\Gamma_\mathrm{inh}/2) \hat{\bi{w}}^\prime]/[\Omega(0)^2 + (\Gamma_\mathrm{inh}/2)^2]^{1/2}$) and lower ($[\Omega(0) \hat{\bi{u}}^\prime - (\Gamma_\mathrm{inh}/2) \hat{\bi{w}}^\prime]/[\Omega(0)^2 + (\Gamma_\mathrm{inh}/2)^2]^{1/2}$) ends are the most detuned ones, at $\Delta = \pm \Gamma_\mathrm{inh}/2$. (b) Scheme of the plane of action of the rAHP control vector when an additional constant phase $\delta$ is included in the field of the rAHP: $\phi_\mathrm{rAHP} = -rt^2/2+\delta$. This plane is rotated an angle $-\delta$ with respect to the plane $u^\prime w^\prime$, where the initial AHP control vector acts ($\phi_\mathrm{AHP} = rt^2/2$).}
\end{center}  
\end{figure}

The ideal initial state right before the application of the ARP-based rephasing sequence is that all the Bloch vectors of the atomic ensemble are aligned in the equatorial plane of the Bloch sphere ($\hat{\bi{u}}^\prime$, for example). We reach this situation in two steps, identical to those described in \cite{Lauro2011}. First, we make use of the 3-level $\Lambda$ system of \tmyag under static magnetic field \cite{deSeze2006} to deplete one of the spin sublevels (noted level $a$) and fully populate the other one (noted level $b$) through optical pumping. This way we set $\bi{B} = \hat{\bi{w}}\prime$. Then, the aim is to rotate that vertical Bloch vector to the equator. The usual way of achieving that is to apply a $\pi/2$ pulse. However, our limited rf power does not allow $\pi/2$ pulses short enough so that their bandwidth would cover $\Gamma_\mathrm{inh}$. Therefore, we turn to an ARP pulse that is interrupted at half its way. The frequency is chirped from $\dot{\phi}(-T/2)$ to $\dot{\phi}(0)$. We will call such pulse an \emph{adiabatic half passage} (AHP). If we neglect the inhomogeneous broadening and if the final frequency of the AHP (or the central frequency of the interrupted ARP) is tuned to the sublevel transition, the AHP (of positive chirp) turns the Bloch vectors from $\hat{\bi{w}}\prime$ to $\hat{\bi{u}}\prime$, which is the desired situation. However, because of the inhomogeneous broadening, the Bloch vectors at the end of the AHP fan out typically from
$[\Omega(0) \hat{\bi{u}}^\prime - (\Gamma_\mathrm{inh}/2) \hat{\bi{w}}^\prime]$
to
$[\Omega(0) \hat{\bi{u}}^\prime + (\Gamma_\mathrm{inh}/2) \hat{\bi{w}}^\prime]$
, as shown in \fref{fig:AHP}(a). This depart from the desired orientation $\hat{\bi{u}}\prime$ will not be relevant, as we will see, since two identical ARPs recover the initial state of \emph{any} Bloch vector ($L=\mathbb{I}$). From this starting point, we execute the rephasing sequence described in \sref{sec:repasingseqmat}. At the end of the sequence, we apply a reversed AHP (rAHP), that is, an AHP where the frequency is chirped in the opposite sense between the same values of the previous AHP. If $L$ satisfies $L=\mathbb{I}$, the rAHP should bring all the Bloch vectors to $\hat{\bi{w}}\prime$, which was the state before the AHP. In fact, this is the main source of information in our experiments: the comparison between the states right before the AHP and right after the rAHP. 

To be more precise about the AHP and rAHP, their phase characteristics for a linear chirp are summarized as

\begin{eqnarray}
\phi_\mathrm{AHP}(t) = r t^2/2 \qquad &\mathrm{and} \qquad -T/2 \leq t \leq 0 \\
\phi_\mathrm{rAHP}(t) = -r t^2/2 \qquad &\mathrm{and} \qquad 0 \leq t \leq T/2 \label{phaserAHP}
\end{eqnarray}


For probing the mean state of the system at any time of the experiment, we measure the intensity of a weak probe pulse tuned to the optical transition from state $a$ to the upper state of the $\Lambda$ system. We assume the probe pulse is weak enough not to alter the distribution of population between the levels. In such a case, the measured intensity $I$ is a function of the population of level $a$ as $I = I_0 \exp(-k \bar{\rho}_{aa})$, with $I_0$ the beam intensity before hitting the sample, $k$ some positive constant determined by the opacity of the sample and $\bar{\rho}_{aa}$ the ensemble average of $\rho_{aa}$. Our conclusions will be drawn from the comparison between $I$ before the AHP, $I_\mathrm{i}$, and $I$ at after the rAHP, $I_\mathrm{f}$.

We performed two experiments. In the first, we aimed at testing the  preservation of the initial superposition state, i.e. $L=\mathbb{I}$ when the two ARPs are identical. In the second, we verified the condition \eref{condition_tau3}: $\tau_3 = \tau_2 - \tau_1$, which gives the rephasing time of the echo.

\subsection{State preservation: Validity of $L=\mathbb{I}$ for identical ARPs}

\begin{figure}
\begin{center}
\includegraphics[width=0.6\textwidth,angle=0]{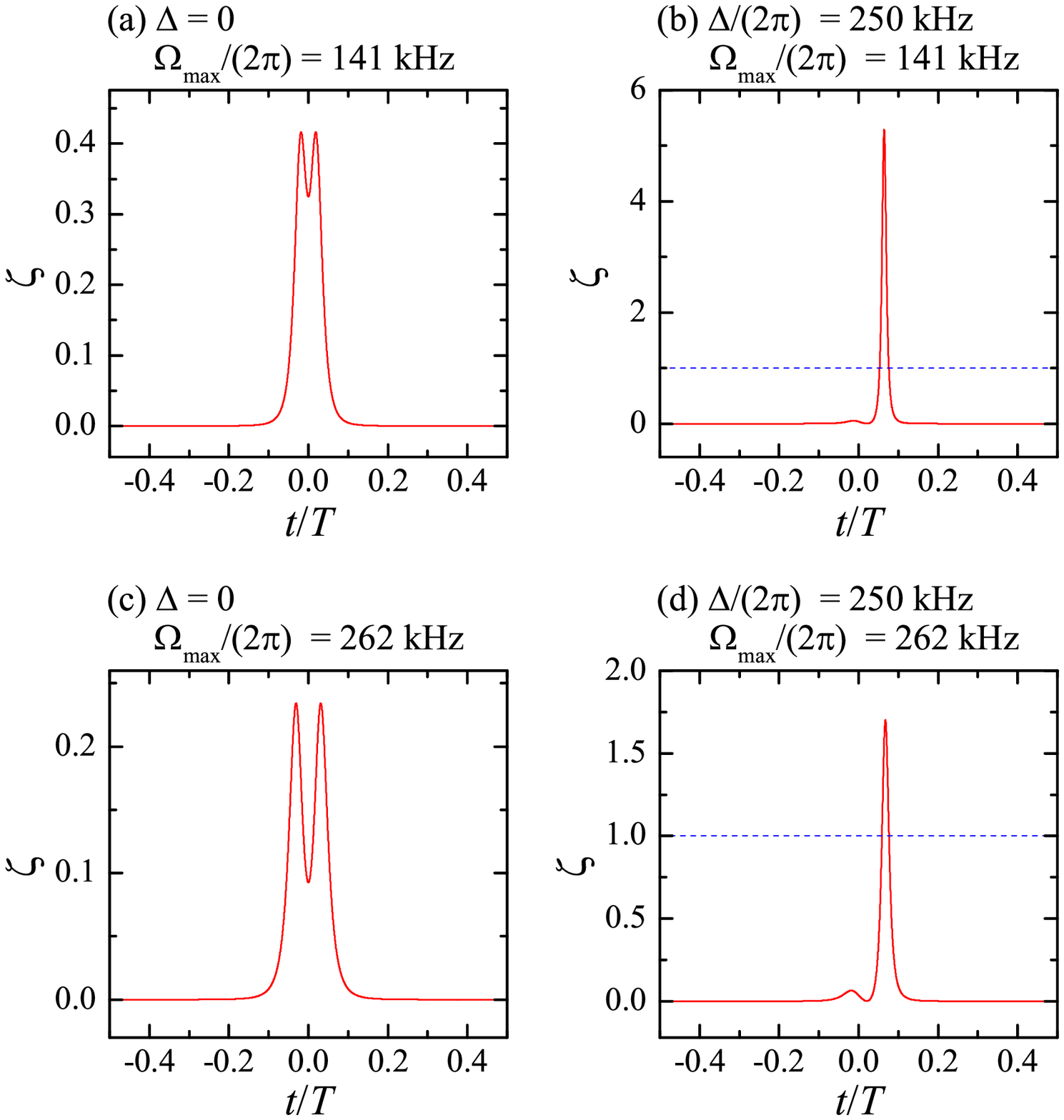}
\caption{\label{fig:adiabaticitycondition} (Colour online) Validity of the adiabatic approximation: $\zeta$ is defined as the l.h.s. of \eref{adiabaticitycondition}, plotted as a function of $t$ for the experimental parameters $\dot{\phi}(\pm T/2)/(2\pi) = \pm 2$ MHz, $T = 100$ $\mu$s, $\phi/(2\pi) = 0.02 (t/\mu\mathrm{s})^2$. Condition \eref{adiabaticitycondition} is not satisfied at every instant of the ARP for frequencies at the edge of $\Gamma_\mathrm{inh}$. (a) $\Delta = 0$, $\Omega_\mathrm{max}/(2\pi) = 141$ kHz: $\zeta < 1$ throughout the ARP. (b) $\Delta/(2\pi) = 250$ $kHz$, $\Omega_\mathrm{max}/(2\pi) = 141$ kHz: $\zeta$ can take values higher than unity. However, $\zeta > 1$ during just $2\%$ of the ARP. (c) $\Delta = 0$, $\Omega_\mathrm{max}/(2\pi) = 262$ kHz. (d) $\Delta/(2\pi) = 250$ $kHz$, $\Omega_\mathrm{max}/(2\pi) = 262$ kHz. The behaviours in (c) and (d) are similar to those in (a) and (b), respectively, except that $\zeta$ takes lower values.}
\end{center}  
\end{figure}

For the particular ARPs used in this experiment, the rf field amplitude at the centre of the circuit resonance yielded a Rabi frequency $\Omega_\mathrm{max}/(2\pi) = 141$ kHz. The rf frequency was chirped linearly from lower to higher frequencies through a range of $4$ MHz during $100$ $\mu$s. As a consequence of its resonance profile, the rf circuit modulates $\Omega(t)$ even though the input rf current amplitude was kept constant throughout the ARP: $\Omega(t)$ follows the circuit profile as the frequency is swept. This ensures fulfilment of the far off-resonance initial and final conditions \eref{condicion180}. To monitor the fulfilment of the adiabatic condition \eref{adiabaticitycondition}, let us define the quantity $\zeta(t)$ as the left-hand side of \eref{adiabaticitycondition}. Using the parameters described above, $\zeta$ is plotted as a function of $t$ in figures \ref{fig:adiabaticitycondition}(a) and \ref{fig:adiabaticitycondition}(b) for two different values of $\Delta$, at the centre and edge of $\Gamma_\mathrm{inh}$. We see that $\zeta <1$ is satisfied by the experimental parameters at the centre of $\Gamma_\mathrm{inh}$ (\fref{fig:adiabaticitycondition}(a)). However, in the case $\Delta/(2\pi) = 250$ kHz (\fref{fig:adiabaticitycondition}(b)), $\zeta <1$ is violated during a short period that amounts to just $2\%$ of the ARP. In any case, we will see that this transient failure of the adiabatic character of the passages for some atomic frequency classes does not invalidate the conclusions drawn from the experiment.

To ensure the  preservation of the initial superposition state, i.e. $L=\mathbb{I}$, quantum process tomography \cite{Nielsen2000} would be the proper experiment to carry out. Here, we will content ourselves with simpler approximate tests.

\begin{figure}
\begin{center}
\includegraphics[width=1.0\textwidth,angle=0]{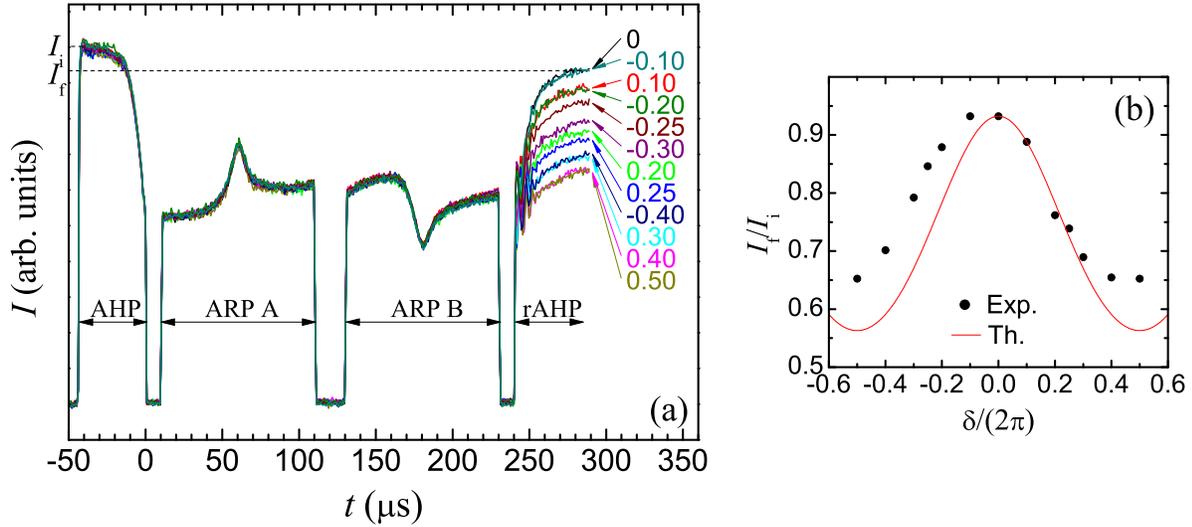}
\caption{\label{fig:phase} (Colour online) (a) Probe intensity profile through a double ARP rephasing sequence. An AHP (rAHP) is applied before (after) the sequence. The parameters for both ARPs are $\Omega_\mathrm{max}/(2\pi) = 141$ kHz and $\phi(t)/(2\pi) = 0.02 (t/\mu \mathrm{s})^2$. Spin $T_2$ in this sample has been measured to be $510$ $\mu$s. Several phases of the rAHP are tested. For each curve, the rAHP phase is set to $-\phi_\mathrm{ARP}+\delta$. The labels indicate $\delta/(2\pi)$. (b) Experimental (symbols) and theoretical (line) ratios between final and initial intensities as a function of the phase of the rAHP.}
\end{center}  
\end{figure}

The first and simplest test is to check if $I_\mathrm{f} = I_\mathrm{i}$. \Fref{fig:phase}(a) shows the intensity profile of the probe beam as the rephasing sequence takes place. We will focus on the curve labelled ``$0$''. At $t=-50$ $\mu$s, the intensity $I_\mathrm{i}$ corresponds to the transparency value (level $a$ is depleted by previous optical pumping). The intensity decreases slowly at a first stage, but close to $t=0$ the intensity fall speeds up. It coincides to the short period where $\bbeta$ is far off the vertical axis. At $t=0$, $I$ attains a value corresponding to $\bar{\rho}_{bb} = \bar{\rho}_{aa}$. After a free evolution time of $10$ $\mu$s (the probe beam is turned off during this and the remaining free evolution intervals), the first ARP starts. The intensity varies slowly during the initial and final regions of the ARP, but it displays a peak at half its way. The origin of this peak is the fact that the Bloch vectors fan out in a circle section contained in the $u^\prime w^\prime$ plane at the end of the AHP, as depicted in \fref{fig:AHP}(a). If all Bloch vectors were aligned at $t=0$, no peak would be observed. Another point is that the intensity should be the same on both sides of the peak (see \fref{fig:taus} as an example). We estimate the intensity difference is due to imperfect matching of the rf frequency to the centre of the inhomogeneous broadening. After a second free evolution time of $20$ $\mu$s, the second ARP is applied. In this case, we observe a dip instead of the peak, of the same origin than that of the latter. The last free evolution interval of $10$ $\mu$s takes place. Finally, the rAHP is applied. The intensity increases to a value $I_\mathrm{f}$ close to the $I_\mathrm{i}$.

The experiment shows that $I_\mathrm{f} < I_\mathrm{i}$. However, we cannot yet conclude that the reason for $I_\mathrm{f}$ lower than expected is imperfect rephasing ($L\neq \mathbb{I}$). As a matter of fact, the observed $I_\mathrm{f}$ is totally compatible with the decrease induced by finite spin coherence lifetime $T_2$. The latter was previously measured to be $\sim 510$ $\mu$s (results not shown). Therefore, as far as this first test is concerned, rephasing is well-achieved by the double ARP sequence. The transient failure of the adiabatic condition for some frequency classes described above does not seem to harm the rephasing efficiency.

The second test consists in monitoring $I_\mathrm{f}$ as the phase of the rAHP is varied from its optimal value given by \eref{phaserAHP}. If a phase $-rt^2/2 + \delta$ is assigned to the linearly chirped rAHP, the control vector during the rAHP will no longer act in the vertical plane $u^\prime w^\prime$. It will act in another vertical plane rotated about $\hat{\bi{w}}^\prime$ an angle $-\delta$ from the original one (see \fref{fig:AHP}(b)). As a consequence, this phase-shifted rAHP will be most suited for driving a Bloch vector that is contained in the same plane at the end of the rephasing sequence. In other words, a rAHP of phase $-rt^2/2 + \delta$ will best fulfil its aim of returning the Bloch vectors to the vertical axis if $L$ is a rotation matrix as in \eref{matrixL} with $\alpha = -\delta$. Thus, this experiment will allow us to check if $\alpha$ is indeed zero in the case of a rephasing sequence with identical ARPs. 

The curves obtained with different values of $\delta$ are exhibited in \fref{fig:phase}(a). The label of each curve corresponds to $\delta/(2\pi)$. We observe that $I_\mathrm{f}$ varies as a function of $\delta$, displaying highest values around $\delta = 0$ and lowest around $\delta = \pi$. This means that when $\delta \simeq 0$ ($\delta \simeq \pi$), the rAHP guides the Bloch vectors mainly to the $\hat{\bi{w}}^\prime$ ($-\hat{\bi{w}}^\prime$) direction.  
The comparison of the experimental ratio $I_\mathrm{f}/I_\mathrm{i}$ with the predictions based on the theory in \sref{sec:repasingseqmat} is presented in \fref{fig:phase}(b). We observe that the maximum is somewhat shifted to the left of the expected value $\delta = 0$. The minimum ratio is also not as low as predicted by the theory. These differences might be due to the slight mismatch between the centre of the frequency chirp range and the centre of $\Gamma_\mathrm{inh}$ and/or to the transient failure of the adiabatic condition for some frequency classes. In any case, it is clear that the overall behaviour is compatible with a sequence matrix close to the one obtained for $L$ with $\alpha=(0.05\pm 0.1)2\pi$.

\subsection{Rephasing time condition: Validity of $\tau_3 = \tau_2-\tau_1$}

For this experiment, the rf circuit was improved to reach a higher Rabi frequency: $\Omega_\mathrm{max}/(2\pi) = 262$ kHz. This allowed us to better satisfy the adiabatic condition \eref{adiabaticitycondition}, as shown in figures \ref{adiabaticitycondition}(c) and \ref{adiabaticitycondition}(d). The remaining parameters of the ARPs were kept identical to those in the previous experiment.

\begin{figure}
\begin{center}
\includegraphics[width=0.7\textwidth,angle=0]{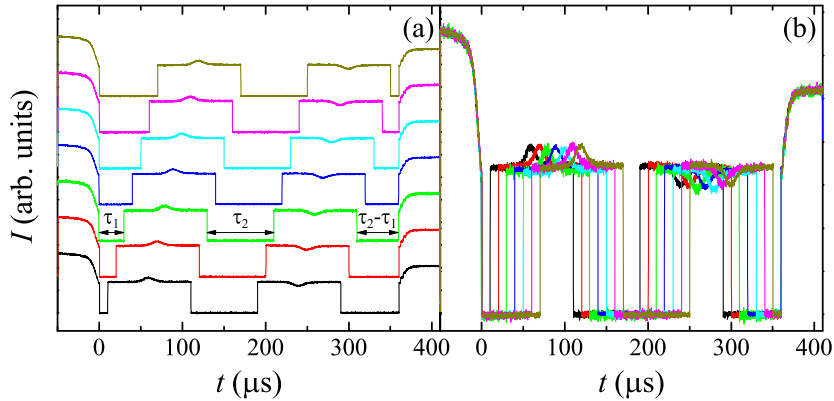}
\caption{\label{fig:taus} (Colour online) ARP rephasing experiments with varying $\tau_1$. $\tau_3$ is adjusted to satisfy $\tau_3=\tau_2-\tau_1$. (a) Stacked curves. (b) Superimposed curves.}
\end{center}  
\end{figure}

We studied the effect on $I_\mathrm{f}$ of varying $\tau_1$ while keeping $\tau_2$ constant. For each rephasing sequence, we adapted $\tau_3$ to fulfil $\tau_3 = \tau_2-\tau_1$ (equation \eref{condition_tau3}). The results are presented in \fref{fig:taus}. Both panels contain the same data (just the way the data are displayed changes). First of all, we observe that the quality of the curves is better than in \fref{fig:phase}(a): the intensity level on both sides of the peak (or dip) are better balanced, which is a signature of better matching of the rf frequency to the centre of the inhomogeneous broadening. Second, the curves look very similar to one another, except, of course, for the positions of the ARPs. The resemblance is better appreciated in \fref{fig:taus}(b), where we clearly see that $I_\mathrm{f}$ is the same for all curves. Here again, the value of $I_\mathrm{f}$ ($<I_\mathrm{i}$) is compatible with the spin decoherence and with the value of $T_2$. Therefore, we conclude that, as far as condition \eref{condition_tau3} is fulfilled, the performance of the rephasing sequence is optimum, independently of the particular values of $\tau_1$ and $\tau_2$.

\section{Summary}

We have thoroughly analysed how rephasing of optical or spin coherences by two ARPs in a double echo scheme works. In the frame of the Bloch sphere formalism, we have theoretically developed a matrix treatment for the sequence, where the resulting matrix $L$ is the product of the matrices associated to each building block. We first derived the matrix for a single ARP, which turned out to be a rotation matrix of an angle $\pi$ about an axis contained in the equator of the Bloch sphere. The particular orientation of this axis within the equator depends both on specific parameters of the ARP and on the atomic frequency. The latter dependence is the reason why one single ARP cannot manage to rephase an inhomogeneously broadened distribution of atomic coherences. A sequence involving two ARPs, however, is able to achieve the rephasing. In our matrix approach, this is understood as $L=\mathbb{I}$, as derived in our calculations as long as some particular conditions are satisfied. We identify these key conditions. One can be satisfied very easily: it is enough to wait a longer time between the two ARPs (time interval $\tau_2$) than between $t=0$ and the first ARP (time interval $\tau_1$). Then, the rephasing takes place a time $\tau_2-\tau_1$ after the end of the second ARP. The second condition implies that the Rabi frequencies and frequency chirps of both ARPs must be identical. The rephasing process can be easily explained with the help of a very simple geometrical interpretation that is drawn from our matrix treatment. 

We have also analysed the capability of the sequence to preserve the initial state phase, assuming the two conditions just mentioned are satisfied. This is of particular importance for quantum memory applications. We have found that phase preservation is assured if the optical or rf fields of both ARPs have exactly the same time-varying phase, meaning that the second ARP field must be a time-shifted copy of the first. This opens a discrepancy between optical and rf rephasing, since the ability of controlling the field phase is different in the rf and optical technologies. In addition, we have provided a quantitative comparison between rephasing by $\pi$ pulses, as in standard echo experiments, or by ARPs. The rephasing efficiency is superior in the latter case.

We have verified experimentally the two rephasing conditions in the rare-earth ion-doped crystal Tm$^{3+}$:YAG. Optically detected NMR experiments evidenced the rephasing capabilities of rf ARP rephasing sequences. If the finite lifetime of the coherence ($T_2$) is taken into account, the experiments compare rather satisfactory to the theoretical predictions.

\ack
This research has been supported by the European Commission through FP7-QuReP(STREP-247743) and FP7-CIPRIS(MC ITN-287252), by the Agence Nationale de la Recherche through ANR-09-BLAN-0333-03 and by the Direction G\'en\'erale de l'Armement.

\section*{References}
\bibliographystyle{iopart-num}
\bibliography{bibliography} 

\end{document}